\documentclass[11pt,a4paper]{article}
\usepackage{amssymb,amsmath,a4wide,graphicx}
\usepackage[numbers,sort&compress]{natbib}
\newcommand{\avg}[1]{E(#1)}
\newcommand{\bavg}[1]{E\big(#1\big)}
\newcommand{\bbavg}[1]{E\bigg(#1\bigg)}

\newcommand{\vek}[1]{\boldsymbol{#1}}
\newcommand{\dd}{\mathrm{d}}
\newcommand{\ee}{\mathrm{e}}
\newcommand{\req}[1]{(\ref{#1})}
\begin{document}
\title{Analysis of Kelly-optimal portfolios}
\author{Paolo Laureti, Mat\'u\v s Medo, Yi-Cheng Zhang}
\date{}
\maketitle
\vspace{-28pt}
\begin{center}
Department of Physics, University of Fribourg,\\
Chemin du Mus\'ee 3, 1700~Fribourg, Switzerland
\end{center}
\vspace{20pt}

\begin{abstract}
\noindent
We investigate the use of Kelly's strategy in the construction
of an optimal portfolio of assets. For lognormally distributed
asset returns, we derive approximate analytical results for the
optimal investment fractions in various settings. We show that
when mean returns and volatilities of the assets are small
and there is no risk-free asset, the Kelly-optimal portfolio
lies on Markowitz Efficient Frontier. Since in the investigated
case the Kelly approach forbids short positions and borrowing,
often only a~small fraction of the available assets is included
in the Kelly-optimal portfolio. This phenomenon, that we call
condensation, is studied analytically in various model scenarios.
\end{abstract}

\section{Introduction}
The construction of an efficient portfolio aims at maximising
the investor's capital, or its return, while minimising the risk
of unfavourable events. This problem has been pioneered by
Markowitz in~\cite{Ma52}, where the Mean-Variance (M-V)
efficient portfolio has been introduced: it minimizes the
portfolio variance for any fixed value of its expected return.
Since this rule can be only justified under somewhat unrealistic
assumptions (namely a quadratic utility function or a normal
distribution of returns, in addition to risk aversion), it
should be considered as a first approximation of the
optimisation process. Later, several optimisation schemes
inspired by Markowitz's work have been
proposed~\cite{Sh64,Per84,KoYa91}. For a recent thorough
overview of the portfolio theory see~\cite{EGBG06}.

A different perspective has been put forward by Kelly
in~\cite{Ke56}, where he shows that the optimal strategy for the
long run can be found by maximising the expected value of the
logarithm of the wealth after one time step. The optimality of
this strategy has long been treated and proven in many different
ways~\cite{Br62,FiWh81,Br00}, according to~\cite{Th00}, it was
successfully used in real financial markets. For an overview of
its continuous time limit see~\cite{PH06}. Recently, the
superiority of typical outcomes to average values has been
discussed from a~different point of view in~\cite{MMZ,MZ}.
Although the Kelly criterion does not employ a utility function,
as pointed out by the author himself, a number of economists
have adopted the point of view of utility theory to evaluate
it~\cite{La59,Sa71,Le73,Ma76}. Various modifications, such as
fractional Kelly strategies~\cite{MZB92} and controlled
drawdowns~\cite{GZ93}, have been proposed to increase security
of the resulting portfolios. A thorough review of the
advantages, drawbacks and modifications of the Kelly criterion
is presented in~\cite{MZ06}. For an exposition of the Kelly
approach in the context of information theory see~\cite{CT06}.

In this paper, we shall discuss the original Kelly strategy in
the framework of a simple stochastic model and without assuming
the existence of a utility function. We will present approximate
analytical results for optimal portfolios in various situations,
as well as numerical solutions and computer simulations. We will
show that, in the limit of small returns and volatilities, when
there is no risk-free asset, the Kelly-optimal portfolio lies on
the Efficient Frontier. Furthermore, we shall analytically study
the conditions under which diversification is no longer
profitable and the optimal portfolio ``condensates'' on a few
assets. Such condensation (or underdiversification) is said to
be typical for the Kelly portfolio~\cite{MZ06} and here we
examine it in various model scenarios. Finally, we will consider
the fluctuations of the logarithm of wealth as a measure of
risk, and compare it with the classic M-V picture.

This paper is organised as follows. After introducing a
multiplicative stochastic model for the dynamics of assets'
prices, we briefly list the main results of the Markowitz
Mean-Variance approach. In Sec.~\ref{GM}, we apply Kelly's
method to our model, analysing the case of one, two and many
risky assets, both with and without additional constraints.
Finally, a combination of the Markowitz Efficient Frontier
with the Kelly strategy is investigated. In the appendix we
explain the approximations used in this paper as well as
a~generalisation of the model to the case of correlated asset
prices.

\subsection{A simple model}
We shall study the portfolio optimisation on a very simple
model which leads to lognormally distributed returns. Consider
$N$ assets, whose prices $p_i(t)$ ($i=1,\dots,N$) undergo
uncorrelated multiplicative random walks
\begin{equation}
\label{model}
p_i(t)=p_i(t-1)\,\ee^{\eta_i (t)}.
\end{equation}
Here the random numbers $\eta_i(t)$ are drawn from Gaussian
density distributions of fixed mean $m_i$ and variance $D_i$,
and are independent of their value at previous time steps. This
model can be easily generalised to the case of non-Gaussian
densities and correlated price variations as it is discussed in
appendix~\ref{app2}; the influence of correlations on the Kelly
portfolio is investigated in~\cite{MYZ09}. We assume that the
investor knows exact values of the parameters $m_i,D_i$---for
the effects of wrong parameter estimates and the details of the
Bayesian parameter-learning process see
\emph{e.g.}~\cite{MSZZ04,MZ06,MPZ08}. We further assume the
existence of a risk-free asset paying zero interest rate.

For the sake of simplicity, we do not include dividends, 
transaction costs and taxes in the model. Hence, the return of
asset $i$ is $R_i(t):=\left[p_i(t)-p_i({t-1})\right]/p_i({t-1})=
\ee^{\eta_i(t)}-1$ is lognormally distributed with the average
$\mu_i:=\avg{R_i}=\exp[m_i+ D_i/2]-1$ and the volatility
$\sigma_i^2:=\avg{(R_i-\mu_i)^2}=(\exp[D_i]-1)\exp[2m_i+D_i]$.
With $E$ we denote averages over the noise $\eta_i(t)$.

A portfolio is determined by the fractions $q_i$ of the total
capital invested in each one of $N$ available assets; the rest
is kept in the~risk-free asset. Since $m_i$ and $D_i$ are fixed,
both the Kelly strategy and the Efficient Frontier use one time
step optimisation and the basic quantity is the wealth after one
time step $W_1$. If we set the initial wealth to 1, $W_1$ has
the form
\begin{equation}
\label{w1}
W_1=1+\sum_{i=1}^N q_iR_i=1+R_P,
\end{equation}
where $R_P:=\sum_{i=1}^N q_iR_i$ is the portfolio return. To
simplify the computation we assume infinite divisibility of the
investment. Thus, the investment fractions $q_i$ are real numbers
and do not need to be rounded.

In the portfolio optimisation, some common constraints are often
imposed and can as well be applied in the present context. For
instance, the non-negativity of the investment fractions
$q_i\geq0$ forbids short positions. The condition
$\sum_{i=1}^N q_i=1$ indicates the absence of a riskless asset
and $\sum_{i=1}^N q_i\leq1$ does not allow the investor to
borrow money.

\subsection{The Mean-Variance approach}
\label{M-V}
The unconstrained maximisation of the expected capital gain
results in the investment of the entire wealth on the asset with
the highest expected return; this strategy is sometimes referred
to as risk neutral. If the investor has a strong aversion to
risk, on the other hand, one might be tempted to simply minimise
the portfolio variance
$\sigma^2_P=\sum_{i=0}^N q_i^2 \sigma^2_i$. This leads to invest
the entire capital on the risk-free asset with no chance to
benefit from asset price movements. The Mean-Variance (MV)
approach is much more reasonable as it allows to compromise
between the gain and the risk. Here we recount basic results of
this standard tool.

With the desired expected return fixed at $\avg{R_P}=\mu_P$, the
constrained minimisation of the portfolio variance $\sigma_P^2$
is performed using the Lagrange function
$\mathcal{L}=\avg{R_P^2}+\gamma\big(\avg{R_P}-\mu_P\big)$ with
a Lagrange multiplier $\gamma$. The resulting optimal fractions
are
\begin{equation}
\label{MV-q}
\hat{q}_i=\mu_P\,\frac{\mu_i}{C_2\sigma_i^2},\quad
\text{where }
C_k=\sum_{j=1}^N\frac{\mu_j^k}{\sigma_j^2}.
\end{equation}
For $\mu_P=0$, $\hat q_i=0$ for all assets. As we increase
$\mu_P$, all optimal fractions $\hat q_i$ grow in a uniform way
and their ratios are preserved. At some value $\mu_P^*$ we reach
$\sum\hat q_i=1$, which means we are investing the entire
capital. Any further increase would require to borrow money,
with Eq.~\req{MV-q} remaining valid as long as the borrowing
rate equals the lending rate (both set to zero here). The
relation between $\sigma_P$ and $\mu_P$ is
\begin{equation}
\label{MV-var}
\sigma_P=\mu_P/\sqrt{C_2}.
\end{equation}
This equation is often referred to as Capital Market Line (CML).
\begin{figure}
\centering
\includegraphics[scale=0.28]{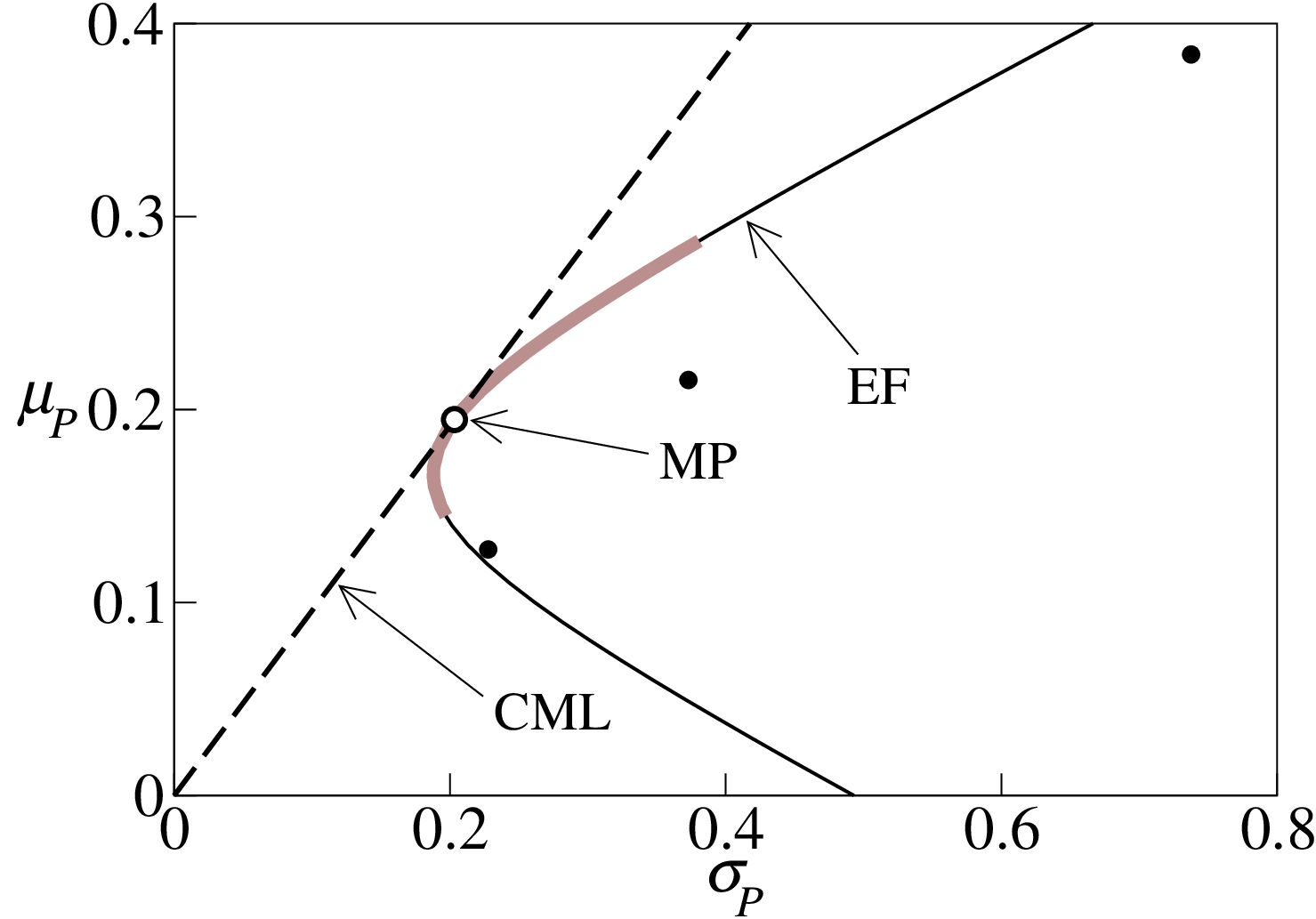}
\caption{The expected return $\mu_P$ versus the standard
deviation of the portfolio $\sigma_P$. The assets are described
by the following parameters: $m_1=0.1$, $D_1=0.04$, $m_2=0.15$,
$D_2=0.09$, $m_3=0.2$, $D_3=0.25$ (in the graph they are shown
as full circles). The dashed line represents the CML from
Eq.~\req{MV-var}, the solid line is Efficient Frontier given by
Eq.~\req{MV-EF}, the tangent point of the two is the Market
Portfolio. The thick part of EF marks the region where all
investment fractions are positive.}
\label{fig-EF}
\end{figure}

If there is no risk-free asset in the market, one has to
introduce the additional constraint
$\sum_{i=1}^N \hat q_i=1$. It follows that
\begin{equation}
\label{MV-EF}
\sigma_P^2=\frac{C_0\mu_P^2-2C_1\mu_P+C_2}{C_0C_2-C_1^2}.
\end{equation}
The functional relation between the optimised $\sigma_P$ and
$\mu_P$ is called Efficient Frontier (EF). Since there is only
one point on the CML where $\sum_i q_i=1$, this line is tangent
to the EF. The results of this section are plotted in
Fig.~\ref{fig-EF} for a particular choice of three available
assets.

\section{The Kelly portfolio}
\label{GM}
When the investor's capital follows a multiplicative process,
after many time steps is its expected value strongly influenced
by rare events and in consequence it is not reasonable to form
a~portfolio by simply maximising $\avg{W(t)}$. The Mean-Variance
approach tries to solve this problem in a straightforward, yet
criticisable way. We support here the idea that an efficient
investment strategy can be found by maximising the investment
growth rate in the long run, which is, under the assumption of
fixed asset properties, equivalent to maximising the logarithm
of the wealth $W_1$ after one time step~\cite{Ke56}. Thus the
key quantity in the construction of a Kelly-optimal portfolio is
$v:=\avg{\ln W_1}$, the average exponential growth rate of the
wealth. We remind that the quantity $\ln W_1$ is not
a~logarithmic utility function.

In~\cite{MZ}, $v$ is optimised in a similar context and the
authors claim that their procedure corresponds to maximising the
median of the distribution of returns. They consider short time
intervals and thus small assets returns. Assuming $R_P\ll1$
(very small portfolio return), they use the approximation
$\ln(1+R_P)\approx R_P-R_P^2/2$ of the logarithm in the
expression of $v$ before maximising it. However, while such an
expansion is only justified for $R_P\ll1$, the maximum of the
resulting function is at $R_P=1$, in contradiction with the
hypothesis. We will develop a~different approximation in the
following.

First, the unconstrained maximisation of $v$ is achieved by
solving the set of equations $\partial v/\partial q_i=0$
($i=1,\dots,N$). After exchanging the order of the derivative
and the average, we obtain the condition
\begin{equation}
\label{max-condition}
\bbavg{\frac{R_i}{1+\sum_j q_j R_j}}=0\qquad (i=1,\dots,N).
\end{equation}
In our case, $R_i$ has a lognormal distribution and to our
knowledge, this set of equations cannot be solved analytically.
With the help of the approximations introduced in the appendix
we shall work out approximative solutions for some particular
cases. We emphasize an important restriction which applies to
all solutions of Eq.~\req{max-condition}. Since returns $R_i$
lie in the range $(-1,\infty)$, when $\sum_{i=1}^N q_i>1$ or
when there is an investment fraction $q_i<0$, there is a~nonzero
probability that $W_1$ is negative and hence $v=\avg{\ln W_1}$
is not well defined. Since Kelly's approach focuses on the long
run, it requires strictly zero probability of getting
bankrupted in one turn. As a consequence, for lognormally
distributed returns any Kelly strategy must obey $q_i\geq 0$
and $\sum_{i=1}^N q_i\leq 1$, i.e. both short selling and
borrowing must be avoided.

\subsection{One risky asset}
Let us begin the reasoning with the case of one risky asset.
We want to find the optimal investment fraction $q$ of the
available wealth. The remaining fraction $1-q$ we keep in cash
at the risk-free interest rate which, without loss of
generality, is set to zero. This problem is described by
Eq.~\req{max-condition} in one dimension; even this simplest
case has no analytical solution. Nevertheless, for a given $D$,
one can ask what is the value $m_<$ for which it becomes
profitable to invest a positive fraction of the investor's
capital in the risky asset. This can be found imposing $q=0$
in Eq.~\req{max-condition}, yielding $m_<=-D/2$. Similarly, the
value $m_>$ for which it becomes profitable to invest the entire
capital can be found by imposing $q=1$, yielding $m_>=D/2$.

\begin{figure}
\centering
\includegraphics[scale=0.28]{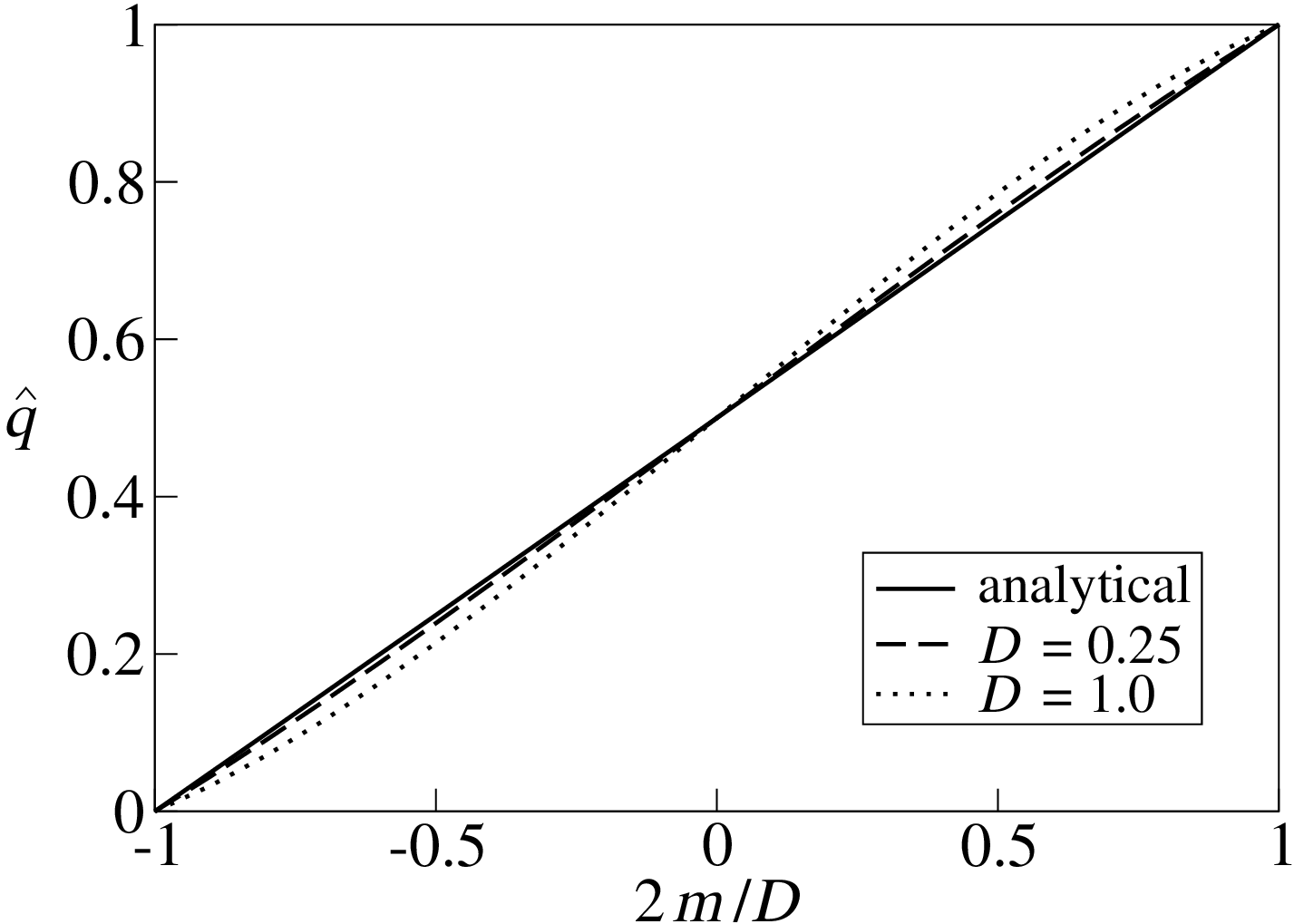}
\caption{The optimal portfolio fraction $\hat q$: a comparison
of the analytical result Eq.~\req{q-analytical} with a numerical
maximisation of $\avg{\ln W_1}$.}
\label{fig-oneasset}
\end{figure}

We shall look for approximate solutions that are valid for small
values of $D$, which is the case treated in appendix~\ref{app1}.
Using approximation Eq.~\req{approximation} in
Eq.~\req{max-condition} gives
$$
\frac{\ee^{m}-1}{1-q+q\ee^{m}}+\frac{D}2\,
\frac{\ee^{m}(1-q-q\ee^{m})}{(1-q+q\ee^{m})^3}=0.
$$
With respect to $q$, this is merely a quadratic equation. Since
the solution is rather long, we first simplify the equation
using $m,D\ll1$ as in Eq.~\req{approx-N}, leading to the result
\begin{equation}
\label{q-analytical}
\hat q=\frac12+\frac mD.
\end{equation}
Since borrowing and short selling are forbidden, for $m<-D/2$ is
$\hat q=0$ and for $m>D/2$ is $\hat q=1$. When asset prices
undergo a multiplicative random walk with lognormal returns,
both $m$ and $D$ scale linearly with the time scale and hence
$\hat q$ does not depend on the length of the time step. Notice
also that substituting $m=\pm D/2$ gives $\hat q=0$ and
$\hat q=1$, in agreement with the bounds we found before by
exact computation. The first order correction to
Eq.~\req{q-analytical} is $m(4m^2-D^2)/4D^2$ which is, for
$m\in[-D/2,D/2]$, of order $O(m)$. The validity of the presented
approximations can be easily tested by a straightforward
numerical maximisation of $\avg{\ln W_1}$. As can be seen in
Fig.~\ref{fig-oneasset}, the numerical results are well
approximated by the analytical formula Eq.~\req{q-analytical}
even for $D=1$.

Notice that, for $m,D\ll 1$, one can approximate
$\mu\approx m+D/2$ and $\sigma^2\approx D$, which makes the
optimal portfolio fraction derived above equal to
$\hat q'=\mu/(\mu^2+\sigma^2)$ obtained in~\cite{MZ}. However,
if we check the accuracy of $\hat q'$, we find a relative error
up to 3\% for $D=0.01$, and for $D=0.25$ we are already far out
of the applicability range with an error around 50\%. Also,
Eq.~\req{q-analytical} is for $m,D\ll 1$ identical to the
classical Merton's result~\cite{Me69} which is derived under
the assumption of continuous-time, non-zero consumption of the
wealth, and a logarithmic utility function. In~\cite{PH06,Br96},
the result $\hat q=\tfrac12+m/D$ is derived for the continuous
time limit of our model: $m,D\to0$, $m/D=\text{const}$.

\subsection{Constrained optimisation}
The optimal portfolio fractions $q_i$ can be derived
from Eq.~\req{max-condition} also for $N>1$. Using the same
approximations as in the single asset case, we obtain the
general formula $\hat q_i=1/2+m_i/D_i$ for $i=1,\dots,N$.
In our case, Kelly's approach forbids short selling and hence
the assets with $\hat q_i<0$ do not enter the optimal portfolio.
Since borrowing is also forbidden, if $\sum_{i=1}^N\hat q_i>1$,
we have to introduce the additional constraint
$\sum_{i=1}^N q_i=1$. This can be done by use of the Lagrange
function $\mathcal{L}(\vek{q},\gamma)=
v+\gamma\,\big(\sum_{i=1}^N q_i-1\big)$ where $\vek{q}$ is the
vector of investment fractions. The optimal portfolio is then
the solution of the set of equations
\begin{equation}
\label{eq-maximum}
\sum_{j=1}^N q_j=1,\quad
\bbavg{\frac{R_i}{1+\sum_{i=1}^N q_iR_i}}+\gamma=0\qquad
(i=1,\dots,N),
\end{equation}
where $R_i=\ee^{\eta_i}-1$. Using the same approximations again,
one obtains the general result
\begin{equation}
\label{qopt-general}
\hat q_i=\frac12+\frac{m_i+\gamma}{D_i}.
\end{equation}
The Lagrange multiplier $\gamma$ is fixed by the condition
$\sum_{j=1}^N \hat q_j=1$. It can occur that even a profitable
asset with $m_i>-D_i/2$ has a negative optimal investment
fraction. Since in our case the Kelly approach forbids short
selling, this asset has to be eliminated from the optimisation
process. In consequence, under some conditions, only a few
assets are included in the resulting optimal portfolio. This
phenomenon, which we call portfolio condensation, we study
closer in sections~\ref{tca} and \ref{tcb}. An alternative
approach to the constrained Kelly-optimal portfolio is provided
by the Kuhn-Tucker equations (see~\cite{CT06}) which, however,
can be shown to be equivalent to~Eq.~\req{eq-maximum}.

Now we can establish an important link to Markowitz's
approach: in the limit $\mu_i,\sigma_i\to0$ the Kelly portfolio
lies on the constrained Efficient Frontier (no short selling
allowed). We shall prove this statement in the following. When
all the assets have small $\mu_i$ and $\sigma_i$, in
Eq.~\req{MV-q} and Eq.~\req{MV-EF} we can approximate
$\mu_i\approx m_i+D_i/2$ and $\sigma_i^2\approx D_i$, leading to
the approximative relation for the Efficient Frontier
\begin{equation}
\label{EF-approx}
\sigma_P^2=\frac{\tilde C_0\mu_P^2-2\tilde C_1\mu_P+\tilde C_2}
{\tilde C_0\tilde C_2-\tilde C_1^2},\quad
\text{where }
\tilde C_k=\sum_{j=1}^N\frac{(m_i+D_i/2)^k}{D_i}.
\end{equation}
For the Kelly portfolio we need to work out a similar
approximation. Using the condition $\sum_i \hat q_i=1$, for
$\gamma$ in Eq.~\req{qopt-general} we obtain
$\gamma=(1-\tilde C_1)/\tilde C_0$. In the relations
$\mu_P=\sum_i q_i\mu_i$ and $\sigma_P^2=\sum_i q_i^2\sigma_i^2$
we use the approximations for $\mu_i,\sigma_i$ introduced above.
After substituting $q_i$ from Eq.~\req{qopt-general}, for the
Kelly optimal portfolio we get
\begin{equation}
\label{xKelly}
\mu_K=\frac{\tilde C_0\tilde C_2-\tilde C_1^2+\tilde C_1}
{\tilde C_0},\quad
\sigma_K^2=\frac{\tilde C_0\tilde C_2-\tilde C_1^2+1}
{\tilde C_0}.
\end{equation}
Both in Eq.~\req{EF-approx} and Eq.~\req{xKelly} we consider
only the assets that have positive investment fractions. Now it
is only a question of simple algebra to show that $\mu_K$ and
$\sigma_K$ given by Eq.~\req{xKelly} fulfill
Eq.~\req{EF-approx}, which completes the proof. Similar, yet
weaker, results can be found in the literature. For instance,
Markowitz states in~\cite{Ma76} that ``on the EF there is
a~point which approximately maximizes $\avg{\ln W_1}$.''

\begin{figure}
\centering
\includegraphics[scale=0.28]{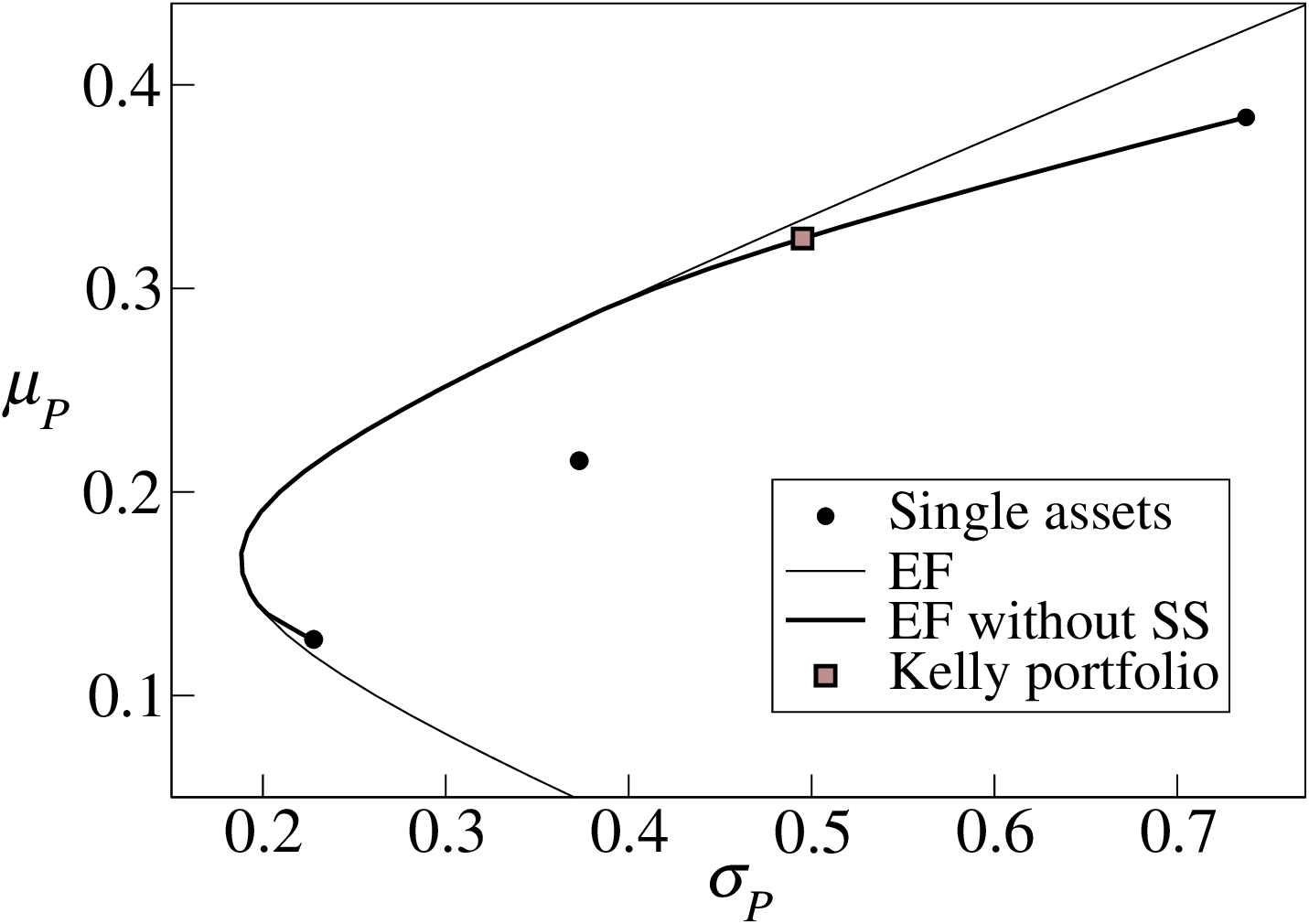}
\caption{The Efficient Frontier (EF, thin solid line), EF
without short selling (bold line), and the Kelly portfolio
(solid square) in a particular case of three assets (asset
parameters as in Fig.~\ref{fig-EF}).}
\label{fig-comparison}
\end{figure}

Obtained results are illustrated in Fig.~\ref{fig-comparison},
where we plot the Efficient Frontier, the constrained Efficient
Frontier, and the Kelly portfolio for the same three assets as
in~Fig.~\ref{fig-EF}. While the original EF is not bounded (for
any $\mu_P$ exists appropriate $\sigma_P$), the constrained EF
starts at the point corresponding to the full investment in the
least profitable asset and ends at the point corresponding to the
most profitable asset. The two lines coincide on a wide range of
$\mu_P$. In agreement with the previous paragraph, the Kelly
portfolio lies close to the constrained EF.

\subsection{Condensation in the two asset case}
\label{tca}
\begin{figure}
\centering
\includegraphics[scale=0.28]{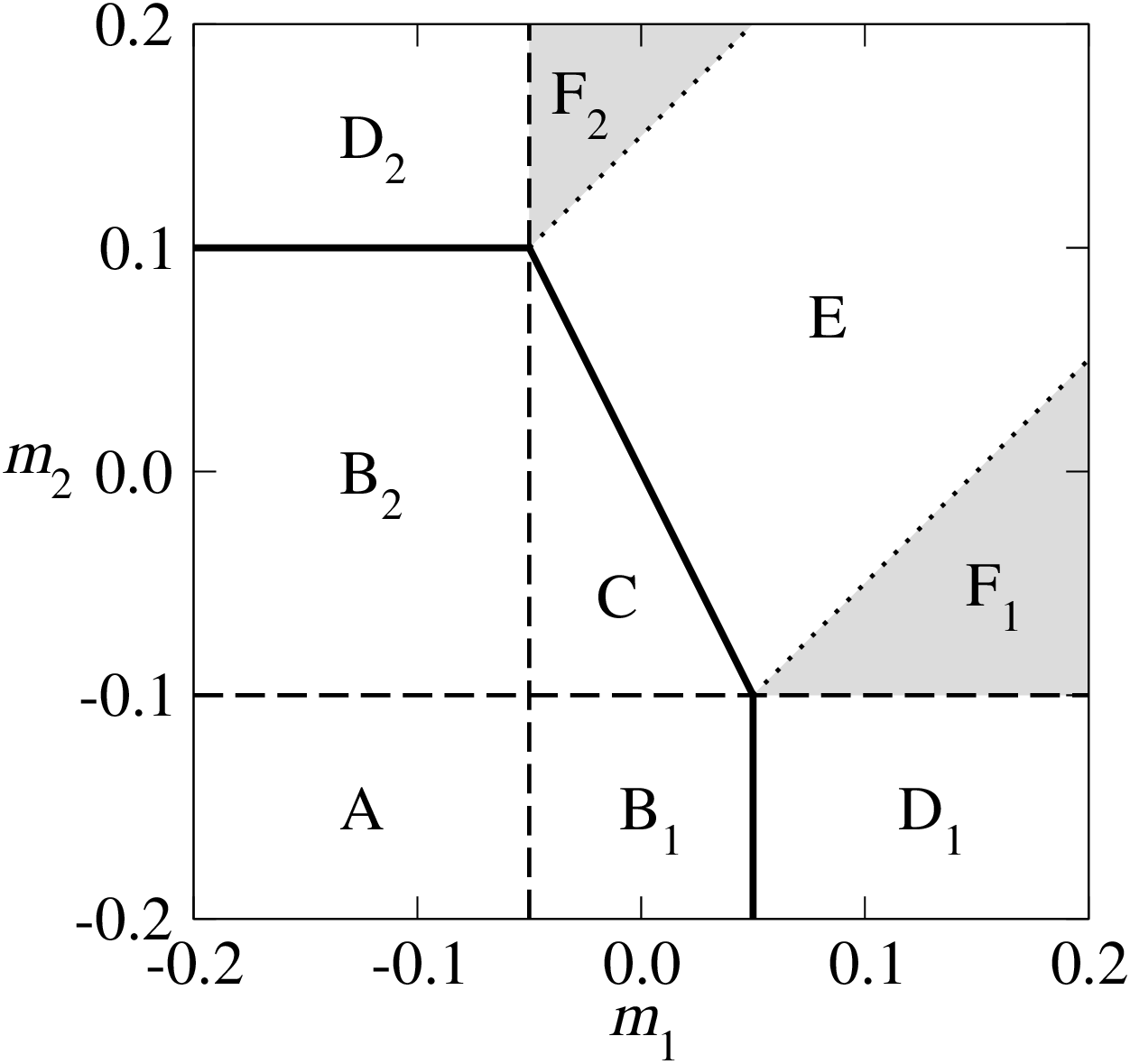}
\caption{The phase diagram of the two-asset system with
$D_1=0.1$ and $D_2=0.2$. In region A the investor is advised to
use only the risk-free asset. In regions B$_1$, B$_2$, and C the
optimal investment is still partially in the risk-free asset. In
regions D$_1$, D$_2$, E, F$_1$, and F$_2$ one should invest
everything in the risky assets. While in regions C and E the
investment is divided between the two assets, in shaded regions
F$_1$ and F$_2$ a nontrivial condensation arises: one is advised
to invest all wealth in one asset although the other one is also
profitable.}
\label{fig-phase_diagram}
\end{figure}

To illustrate the condensation phenomenon we focus on a simple
case here: two risky assets plus a risk-free one, borrowing and
short selling forbidden. As we have already seen, without
constraints $\hat q_i=1/2+m_i/D_i$. Therefore, when $m_i<-D_i/2$,
$\hat q_i$ is negative and due to forbidden short selling, asset
$i$ drops out of the optimal portfolio. In
Fig.~\ref{fig-phase_diagram} this threshold is shown for $i=1,2$
by dashed lines. In the lower-left corner (A) we have the region
where both assets are unprofitable and the optimal strategy
prescribes a fully riskless investment.

When the results of the unconstrained optimisation sum up to one
($\hat q_1+\hat q_2=1$), we are advised to invest all our
wealth in the risky assets. If both assets are profitable, this
occurs when $m_1/D_1+m_2/D_2=0$. When only asset $i$ is
profitable, we should invest the entire capital on it only when
$m_i$ equals at least $D_i/2$. In Fig.~\ref{fig-phase_diagram}
these results are shown as a thick solid curve.

Since borrowing is not allowed, in the region above the solid
line constrained optimisation has to be used. The condensation
to one of the two assets arises when the optimal fractions
$(\hat q_1,\hat q_2)$ are either $(1,0)$ or $(0,1)$; we can find
the values $m_1'$ and $m_1''$ when this happens. By eliminating
$\gamma$ from Eq.~\req{eq-maximum} and substituting $q_1=1$ and
$q_2=0$ we obtain the condition for the condensation on asset
$1$: $\bavg{(\ee^{\eta_1}-1)/\ee^{\eta_1}}=
\bavg{(\ee^{\eta_2}-1)/\ee^{\eta_1}}$. This can be solved
analytically, yielding
\begin{equation}
\label{condensation}
m_1'=m_2+\frac{D_1+D_2}2.
\end{equation}
This equation holds with interchanged indices for the
condensation on asset $2$, thus $m_1''=m_2-(D_1+D_2)/2$.
Finally, for $m_1''<m_1<m_1'$ the optimal portfolio contains
both assets. The crossover values $m_1'$ and $m_2'$ are shown in
Fig.~\ref{fig-phase_diagram} as dotted lines. They delimit the
region where the portfolio condensates to only one of two
profitable assets. A complete ``phase diagram'' of the optimal
investment in the two assets case is presented in
Fig.~\ref{fig-phase_diagram} for a particular choice of the
assets' variances. Interestingly, growth-rate optimising
strategies have their importance also in evolutionary
biology~\cite{YoJa96} where a similar condensation phenomenon
has been observed when studying evolution in an uncertain
environment~\cite{BeLa04}.

\subsection{Many assets with equal volatility}
\label{tcb}
We investigate here the case of an arbitrary large number $N$ of
available assets where Kelly's approach, which forbids borrowing
and short selling, gives rise to a portfolio condensation. While
the optimal portfolio fractions are given by
Eq.~\req{qopt-general}, to find which assets are included in the
optimal portfolio is a hard combinatorial task. To obtain
analytical results, we simplify the problem by assuming that the
variances of all assets are equal, $D_i=D$ ($i=1,\dots,N$). The
number of assets contained in the optimal portfolio is labeled
as $M$ and the assets are sorted in order of decreasing $m_i$
($m_1>m_2>\dots>m_N$).

If the unconstrained optimisation does not violate forbidden
borrowing, then profitability of an asset (i.e., $m_i>-D/2$) is
the only criterion for including it in the optimal portfolio.
When the constrained optimisation is necessary, the optimal
portfolio is formed by starting from the most profitable asset
$m_1$, and adding the others one by one until the last added
asset has a nonpositive optimal fraction $q_{M+1}\le 0$. Summing
Eq.~\req{qopt-general} from $1$ to $M$, we can write $\gamma$ as
$\gamma(M)=D(M^{-1}-1/2)-\tfrac1M\sum_1^Mm_i$. For a given
realisation of $\{m_i\}$, we can obtain the resulting portfolio
size by finding the largest $M$ that satisfies $q_M>0$, which
leads to
\begin{equation}
\label{optimal-special}
m_M+\frac{D}{M}>\frac1M\sum_{i=1}^M m_i.
\end{equation}
This relation tells us how many assets we should invest on, once
their expected growths and volatility are known. Notice that for
$M=2$ and $D_1=D_2=D$, this result is consistent with that of
Eq.~\req{condensation} where a special case of the condensation
on two assets is described.

Let us follow now a statistical approach. If all $m_i$ are drawn
from a given distribution $f(m)$, the value of $M$ depends on
the current realisation. The characteristic behaviour of the
system can be found by taking the average over all possible
realizations and replacing $m_i$ by $\overline{m}_i$. The
resulting typical portfolio size $M_T$ captures this behaviour
and depends on the distribution $f(m)$ and on the number of
available assets $N$.

\subsubsection{Uniform distribution of $m$}
Let us first analyse the case of a uniform distribution of $m_i$
within the range $[a,b]$. First we assume that all assets are
profitable, i.e. $a+D/2>0$. For $m_i$ are sorted in decreasing
order, one can show $\overline{m}_i=b-(b-a)i/(N+1)$. Since
$\overline{m}_i$ declines with $i$ linearly, according to
Eq.~\req{qopt-general} so does $\overline{\hat q_i}$.
Substituting $\overline{m}_i$ for $m_i$ in
Eq.~\req{optimal-special} and replacing $>$ with $=$, we can
estimate the typical number of assets in the optimal portfolio
$M_T$. Assuming $M_T\gg1$, the solution has the simple form
$M_T=\sqrt{2ND/(b-a)}$.

\begin{figure}
\centering
\includegraphics[scale=0.28]{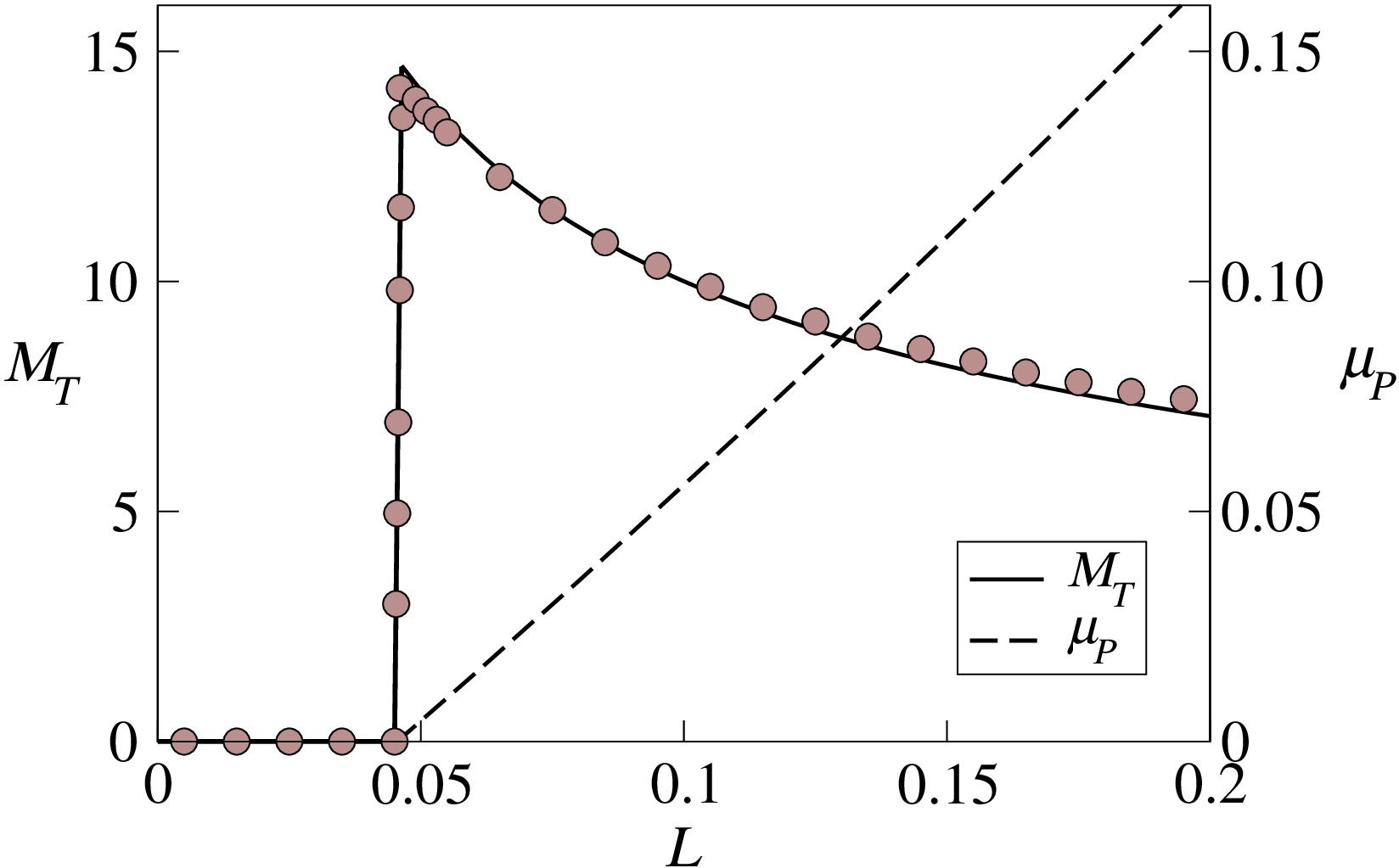}\quad
\includegraphics[scale=0.28]{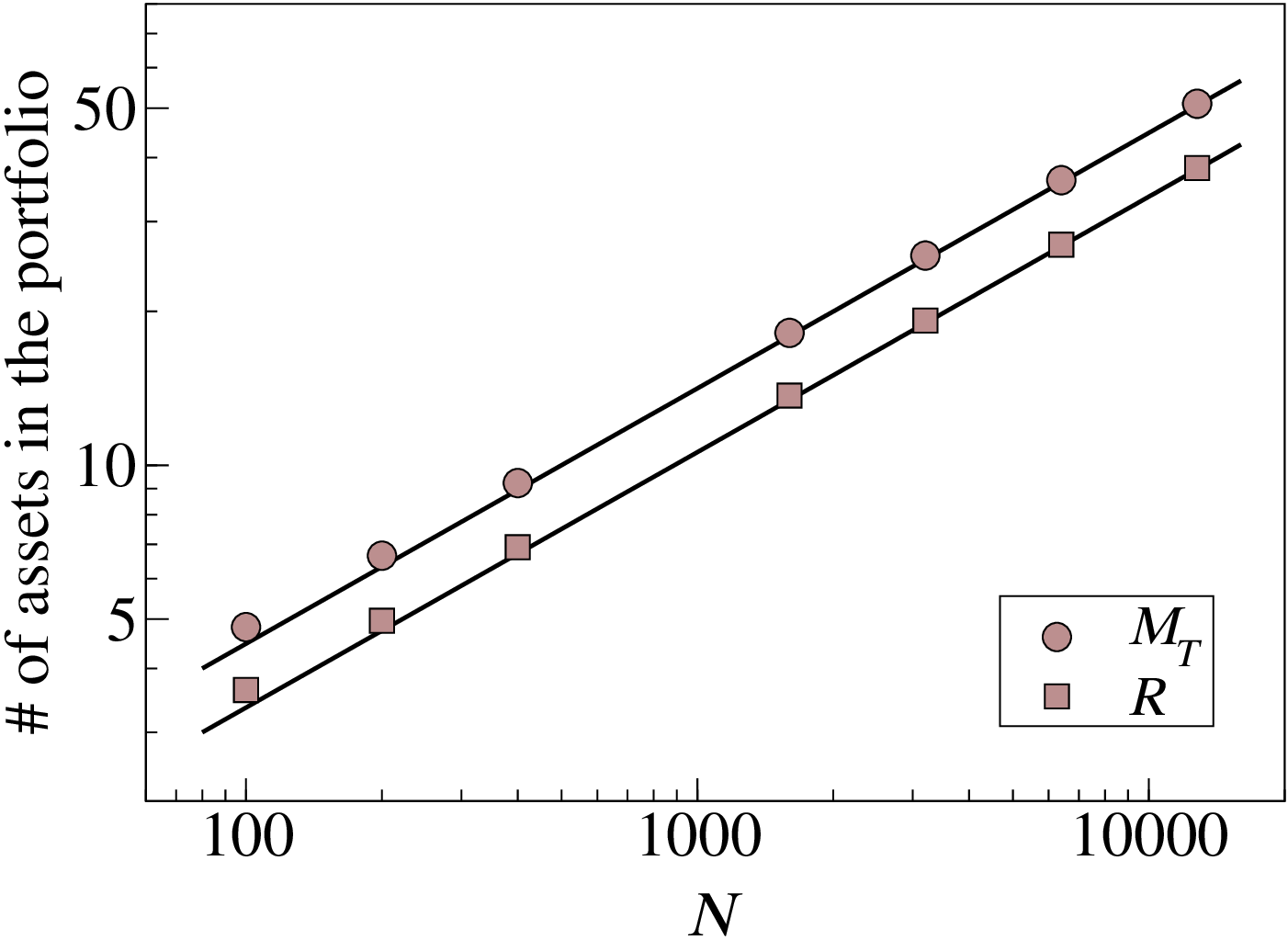}
\caption{{\it Left}: In the case with $m_i$ uniformly
distributed in $[x-L,x+L]$ we plot the size of the optimal
portfolio $M_T$ (solid line represents analytical
result Eq.~\req{uniform-general}, symbols are numerical results
averaged over $10\,000$ repetitions). With the dashed line the
portfolio return $\mu_P$ (measured in percents) is shown. The
parameters are $N=1\,000$, $D=0.01$, $x=-0.05$.
{\it Right}: In the case of $N$ assets with $D=0.01$ and $m_i$
uniformly distributed in $[0,0.1]$, we plot the average size of
the optimal portfolio $M_T$, and the inverse participation ratio
$\mathcal{R}$. Solid lines are the analytical solutions reported
in Eqs.~\req{uniform-general} and~\req{part-rat}, symbols stand
for numerical results.}
\label{fig-uniform}
\end{figure}

We are now able to generalise this result to the case where not
all assets are profitable, i.e. $m_i+D/2<0$ for some $i$'s. In
the extreme case $b+D/2<0$ and all assets are unprofitable,
leading to $M=M_T=0$. The opposite extreme is realized for
$m_{M_T+1}+D/2>0$ which falls in the previously treated case
because the number of profitable assets is larger than $M_T$. In
the intermediate region, only the assets with $m_i>-D/2$ are
profitable and enter the optimal portfolio. On average, they are
$N(b+D/2)/(b-a)$. All together we have the formula
\begin{equation}
\label{uniform-general}
M_T=\begin{cases}
0 & (b+D/2<0),\\
N\,\frac{b+D/2}{b-a} &
(b+D/2>0,\ \overline{m}_{M_T}+D/2<0),\\
\sqrt{\frac{2ND}{b-a}} & (\overline{m}_{M_T}+D/2>0).
\end{cases}
\end{equation}
In Fig.~\ref{fig-uniform}, left, an illustration of a particular
system ($a=x-L$, $b=x+L$, $x=-0.05$, motivated by~\cite{AC}) is
shown. We plot $M_T$ and $\mu_P$ as functions of $L$. When $L$
is small, all available assets are unprofitable and the optimal
strategy is to keep the entire capital at the risk-free rate. As
soon as the first profitable assets are added to the system, the
optimal portfolio includes all of them, until it saturates at the
value $\sqrt{ND/L}$ (in Eq.~\req{uniform-general} we substitute
$b-a=2L$). A further increase of $L$ widens the distribution of
$m$ and enlarges the gaps between profitable assets. It becomes,
as a consequence, more rewarding to drop the worse ones and
$M_T$ decreases. The analytical solution, displayed in
Fig.~\ref{fig-uniform} as a solid line, is in a good agreement
with the numerical results (shown as symbols). Although no
single-asset portfolio arises in this case, the relative
portfolio size is $M_T/N\sim1/\sqrt{N}$ and hence in the large
$N$ limit, the optimal portfolio condensates to a small fraction
of all available assets.

A more flexible measure of the level of condensation is the
\emph{inverse participation ratio}, defined as
$\mathcal{R}=1/\sum_{i=1}^N q_i^2$. It estimates the effective
number of assets in the portfolio: when all investment fractions
are equal, $\mathcal{R}=N$, while when one asset covers 99\% of
the portfolio, $\mathcal{R}\approx1$. Concerning the typical
case, using Eq.~\req{qopt-general} we can write
$\overline{q}_i=A-B i$, $B=(b-a)/[D(N+1)]$, the detailed form of
$A$ is not needed for the solution. We assume that the number
of profitable assets is larger than the typical size of the
optimal portfolio $M_T$. Consequently, passing from $i=1$ to
$i=M_T$, $q_i$ decreases linearly to zero and we can use the
identity $\sum_{i=1}^{M_T} (A-Bi)^2=\sum_{i=1}^{M_T} (Bi)^2$ to
obtain
\begin{equation}
\label{part-rat}
\mathcal{R}=\big[B^2M_T(M_T+1)(2M_T+1)/6\big]^{-1}\approx
\big[B^2M_T^3/3\big]^{-1}\approx\tfrac34M_T.
\end{equation}
In the last step we used Eq.~\req{uniform-general} for the
typical size of the condensed optimal portfolio. We see that the
uniform distribution of $m_i$ leads to the inverse participation
ratio proportional to the number of assets in the portfolio. In
the right graph of Fig.~\ref{fig-uniform}, Eq.~\req{part-rat} is
shown to match the numerical solution (based on
Eq.~\req{optimal-special}) for various numbers of available
assets.

\subsubsection{Power-law distribution of $m$}
Now we treat the case of a distribution $f(m)$ that has a
power-law tail: $f(m)=Cm^{-\alpha-1}$ for $m>m_{\min}$. As long
as $M\ll N$, the properties of the assets included in the
optimal portfolio are driven by the tail of $f(m)$. In
consequence, the detailed form of $f(m)$ for $m<m_{\min}$ is not
important here. We assume that only a fraction $r$ of all assets
falls in the region $m>m_{\min}$.

\begin{figure}
\centering
\includegraphics[scale=0.28]{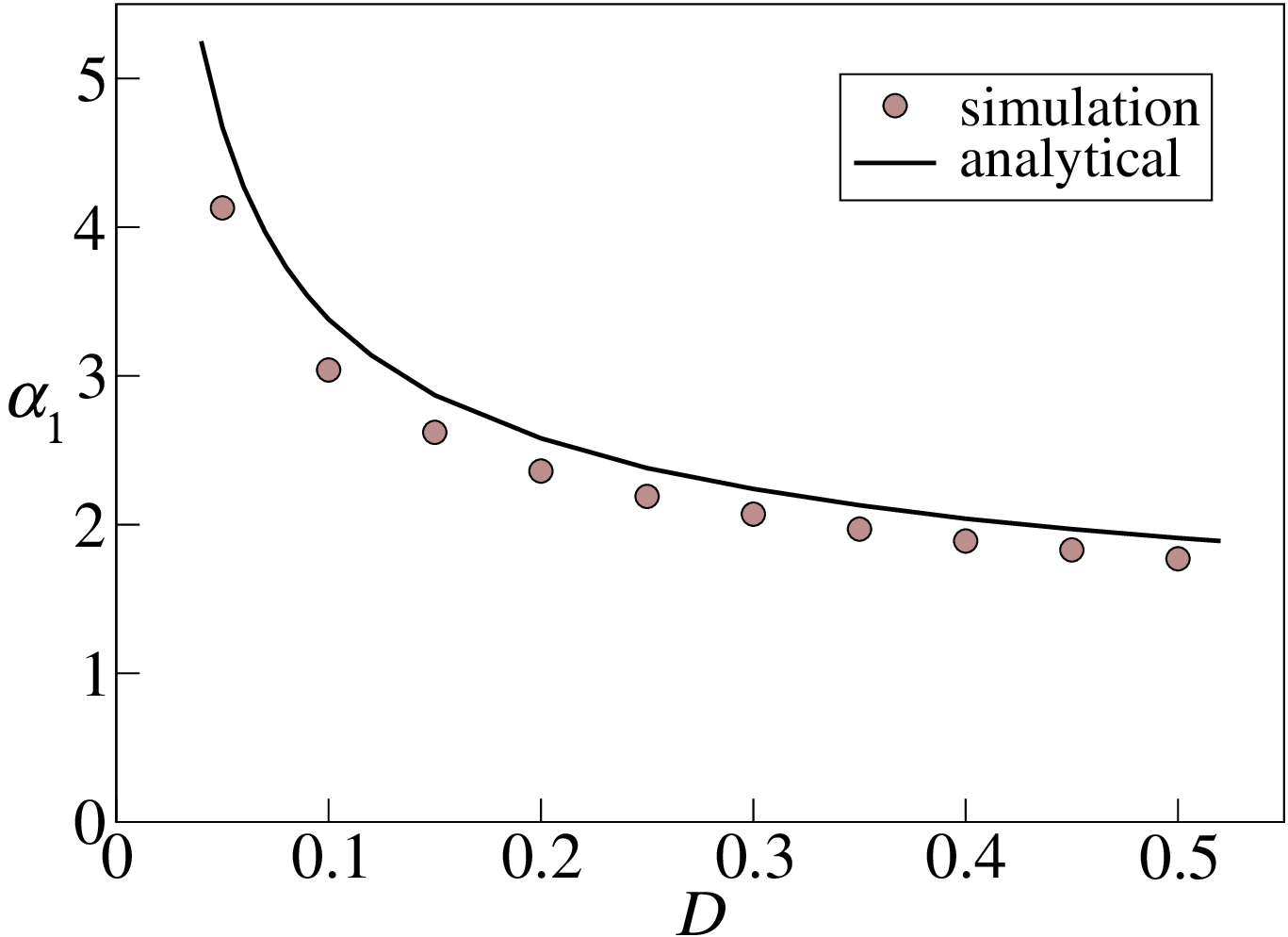}
\caption{Values of the power law exponent $\alpha_1$ at which
condensation to one asset arises for $N=1\,000$, $r=0.1$, and
$m_{\min}=0.1$. We compare the semi-analytical result obtained
by solving $\tilde m_1-\tilde m_2=D$ (shown as the solid line)
with a numerical simulation of the system (shown as filled
circles). Below the line, the optimal portfolio typically
contains only one asset.}
\label{fig-power-law}
\end{figure}

Instead of seeking the typical portfolio size $M_T$, we shall
limit ourselves to finding the conditions when a condensation on
one asset arises. With this aim in mind, we put $M=2$ in
Eq.~\req{optimal-special}, obtaining the equation
$m_1-m_2=D$. When $m_1-m_2>D$, only asset 1 is included in the
optimal portfolio. By replacing $m_1$ and $m_2$ with their
medians $\tilde m_1$ and $\tilde m_2$, one obtains an
approximate condition for a system where such a condensation
typically exists. Using \emph{order statistics}~\cite{DN03} we
find the following expressions for the medians:
$\tilde m_1=m_{\min}(Nr/\ln2)^{1/\alpha}$,
$\tilde m_2\doteq m_{\min}(Nr/1.68)^{1/\alpha}$. The equation
$\tilde m_1-\tilde m_2=D$ thus achieved can be solved
numerically with respect to $\alpha$. In this way we find the
value $\alpha_1$ below which the optimal portfolio typically
contains only the most profitable asset.
In~Fig.~\ref{fig-power-law} we plot the result as a function of
$D$. For comparison, the outcomes from a purely numerical
investigation of the equation $P(m_1-m_2>D)=0.5$ are also shown
as filled circles. Our approximate condition has the same
qualitative behaviour as the simulation, showing that the use of
median gives us a good notion of the optimal portfolio behaviour.

\section{Efficient frontiers}
\label{EF}
Markowitz's Efficient Frontier is the line where efficient
portfolios are supposed to lie in the Mean-Variance picture.
Here we would like to follow the same procedure, using typical
instead of average quantities. To capture a typical case, we
replace the portfolio return $R_P$ by $\ln W_1$ in all averages
of Sec.~\ref{M-V}. According to the formula
$\avg{(x-\avg{x})^2}=\avg{x^2}-\avg{x}^2$ we can minimise
$\avg{(\ln W_1)^2}-\avg{\ln W_1}^2$ instead of
$\avg{(\ln W_1-\avg{\ln W_1})^2}$. With the constraints
$\avg{\ln W_1}=v_P$ and $\sum_{i=1}^N q_i=1$, the Lagrange
function has the form
\begin{equation}
\label{lagrange}
\mathcal{L}=\bavg{(\ln W_1)^2}+
\gamma_1\big(\avg{\ln W_1}-v_P\big)+
\gamma_2\big(\sum_{i=1}^N q_i-1\big).
\end{equation}
Its analytical maximisation leads to complicated equations and
thus it is convenient to investigate the system numerically; we
do so in the particular case of three assets used in
Fig.~\ref{fig-EF}. Due to the two constraints there is
effectively only one degree of freedom for the minimisation of
$\avg{(\ln W_1)^2}$ and the numerical procedure may be
straightforward.

\begin{figure}
\centering
\includegraphics[scale=0.28]{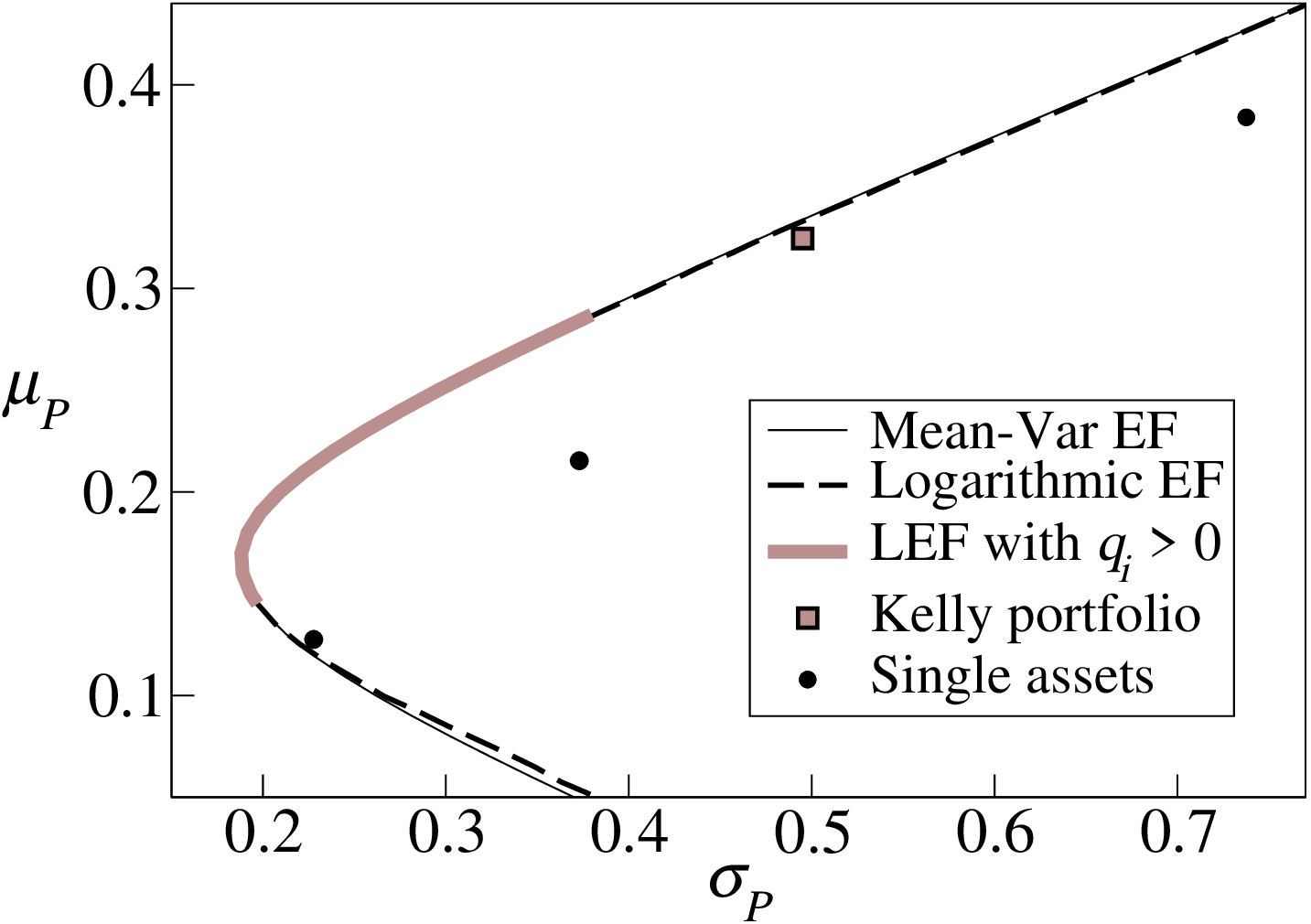}
\caption{Comparison of two efficient frontiers. The solid line
is the classic EF given by Eq.~\req{MV-EF}. The dashed line is
obtained by fixing $\avg{\ln W_1}$ and by numerically minimising
fluctuations of $\ln W_1$ around this value. This we call LEF,
its portion with positive $q_i$ values is highlighted in thick
gray. The small filled circles represent the three individual
assets that compose the system. The filled square is the
Kelly-optimal portfolio (no borrowing, no short selling).}
\label{EFcomp}
\end{figure}

For the resulting portfolios we can compute expected returns and
variances which allows us to add this ``Logarithmic Efficient
Frontier'' (LEF) to the $\sigma_P$-$\mu_P$ plane depicted in
Fig.~\ref{fig-EF}. The result is shown in Fig.~\ref{EFcomp},
where the solid line is again Markowitz's EF. Solid circles
correspond to the three individual assets. The dashed line
represents LEF (obtained by the numerical optimisation described
above) and the thick gray curve is the region where both EF and
LEF consist only of positive portfolio fractions. The solid
square represents the Kelly portfolio as follows from
Eq.~\req{qopt-general} (again, short selling and borrowing are
forbidden). We see that EF and LEF are close to each other and
thus from the practical point of view they do not differ.

Finally, let us discuss a useful simplification which allows us,
in some cases, to reduce the time-consuming numerical
computations. By differentiating Eq.~\req{lagrange} we obtain
the condition for the optimal portfolio fractions
\begin{equation}
\label{newconditions}
\bbavg{\frac{2\ln W_1}{W_1}\,R_i}+
\gamma_1\,\bbavg{\frac{R_i}{W_1}}+\gamma_2=0\qquad
(i=1,\dots,N).
\end{equation}
When the parameters of all assets fulfill the condition
$m_i,D_i\ll 1$, we can use approximations for $\avg{R_i/W_1}$
introduced in Appendix~\ref{app1}. The first term can be
evaluated more precisely using
$\ln W_1=\ln(1+R_P)\approx R_P$. Hence
$\avg{{R_i\ln W_1}/{W_1}}\approx q_i\,\avg{R_i^2}+
\sum_{j\neq i}q_j\,\avg{R_i}\avg{R_j}$,
where $R_i=\ee^{\eta_i}-1$ is the return of asset $i$.
Furthermore, for $m_i,D_i\ll 1$ we have
$\avg{R_i}\approx m_i+D_i/2$ and $\avg{R_i^2}\approx D_i$. As a
result we obtain the equations
\begin{equation}
\label{LEF}
2q_iD_i+2(m_i+D_i/2)\sum_{j\neq i}q_j(m_j+D_j/2)+
\gamma_1\bigg[m_i+\frac{D_i}2\,(1-2q_i)\bigg]+\gamma_2=0,
\end{equation}
where $i=1,\dots,N$ and the values of $\gamma_1$ and $\gamma_2$
are fixed by the constraints $\sum_{i=1}^N q_i(m_i+D_i/2)=v_P$,
$\sum_{i=1}^N\hat q_i=1$. This set of $N+2$ nonlinear equations
allows us to approximately find Logarithmic Efficient Frontier.
In comparison with a straightforward numerical maximisation of
Eq.~\req{lagrange} (involving numerical integration of
$\avg{\ln W_1}$ and $\avg{(\ln W_1)^2}$), a substantial saving
of computational costs is achieved.

\section{Concluding remarks}
In this work we investigated the Kelly optimisation strategy in
the framework of a simple stochastic model for asset prices. We
derived a highly accurate approximate analytical formula for the
optimal portfolio fractions. We proved that in the limit of
small returns and volatilities of the assets, the constrained
Kelly-optimal portfolio lies on the Efficient Frontier. Based on
the obtained analytical results, we proposed a simple algorithm
for the construction of the optimal portfolio in the constrained
case. We showed that since in the investigated case of lognormal
returns, Kelly's approach forbids short positions and borrowing,
only a part of the available assets is included in the optimal
portfolio. In some cases the size of the optimal portfolio is
much smaller than the number of available assets---we say that
a~portfolio condensation arises. In particular, when the
distribution of the mean asset returns is wide, there is a high
probability that only the most profitable asset is included in
the Kelly-optimal portfolio.

The Mean-Variance analysis is a well-established approach to the
portfolio optimisation. We modified this method by replacing the
averages $\avg{W_1}$ and $\avg{W_1^2}$ with the logarithm-related
quantities $\avg{\ln W_1}$ and $\avg{(\ln W_1)^2}$. These are
less affected by rare events and allow to capture the typical
behaviour of the system. As a matter of fact, the difference
between the traditional M-V approach and the modification
proposed here is very small and does not justify the additional
complexity thus induced.

\section{Acknowledgement}
We acknowledge the partial support from Swiss National Science
Foundation (project 205120-113842) as well as STIPCO (European
exchange program). We appreciate early collaboration with Dr.
Andrea Capocci in this research, stimulating remarks from
Damien Challet in late stages of the paper preparation, and
helpful comments of our anonymous reviewers.

\appendix
\section{Main approximations}
\label{app1}
Our aim is to approximate expressions of the type
$\avg{g(\eta)}$, where $\eta$ follows a normal
distribution $f(\eta)$ with the mean $m$ and the variance $D$.
For small values of $D$, this distribution is sharply peaked and
an approximate solution can be found expanding $g(\eta)$ around
this $m$. This expansion has the following effective form
\begin{equation}
\label{expansion}
g(\eta)\stackrel{\text{ef.}}{=}g(m)+
\frac12(\eta-m)^2g^{(2)}(m)+
\frac1{24}(\eta-m)^4g^{(4)}(x),
\end{equation}
for some $x\in[m,\eta]$. Here we dropped the terms proportional
to $(\eta-m)^k$ with an odd exponent $k$, for they vanish after
the averaging. If we take only the first two terms into account,
we obtain
\begin{equation}
\label{approximation}
\bavg{g(\eta)}=
\int_{-\infty}^{\infty}g(\eta)f(\eta)\,\dd\eta\approx
g(m)+\frac D2\,g^{(2)}(m).
\end{equation}
This approximation is valid when the following term of the Taylor
series brings a negligible contribution $\Delta$. We can
estimate it in the following way ($x\in[m,\eta]$)
$$
\Delta=\int_{-\infty}^{\infty}\frac{(\eta-m)^4}{24}
g^{(4)}(x)f(\eta)\,\dd\eta\lesssim
\int_{-\infty}^{\infty}\frac{(\eta-m)^4}{24}\,M
f(\eta)\,\dd\eta=\frac{MD^2}8.
$$
Here by $M$ we label the maximum of $\lvert g^{(4)}(\eta)\rvert$
in the region $\mathcal{X}$ where $f(\eta)$ differs from zero
considerably, e.g. $\mathcal{X}=[m-2D,m+2D]$. Since $g(x)$
has no singular points in a wide neighbourhood of $m$,
its fourth derivative is a bounded and well-behaved function.
Thus $M$ is finite and $\Delta$ vanish when $D$ is small.

In particular, in this work we deal with functions of the form
$g(\eta_i)=(\ee^{\eta}-1)/[1+q(\ee^{\eta}-1)]$. If we use
Eq.~\req{approximation} with this $g(\eta)$, approximate
$1+q(\ee^m-1)$ in the resulting denominators by $1$, $\ee^{m}$
by $1$, and $\ee^{m}-1$ by $m$, we are left with
\begin{equation}
\label{approx-N}
\bavg{g(\eta)}\approx m+D\big(1-2q\big)/2.
\end{equation}
We widely use approximations of this kind to obtain the leading
terms for the optimal portfolio fractions in this paper.

\section{Procedure for correlated asset prices}
\label{app2}
So far we have considered uncorrelated asset prices, undergoing
the geometric Brownian motion of Eq.~\req{model}. Obviously,
this is an idealised model and real asset prices exhibit various
kinds of correlations. In order to treat correlated prices we
employ the covariance matrix $\mathsf{S}$ to characterise the
second moment of the stochastic terms
$\avg{(\eta_i-m_i)(\eta_j-m_j)}=S_{ij}$. The uncorrelated case
can be recovered with the substitution
$S_{ij}=\delta_{ij}D_{i}$.

Again, we would like to find an approximation of the term
$\bavg{g(\vek{\eta})}\equiv
\int g(\vek{\eta})f(\vek{\eta})\,\dd\vek{\eta}$. Here
$f(\vek{\eta})$ is the probability distribution of $\vek{\eta}$
and $g(\vek{\eta})$ is the function of interest. Notice that the
correlations impose the use of vector forms for all the
quantities of interest. The Taylor expansion of $g(\vek{\eta})$
around $\vek{m}$, Eq.~\req{expansion} in the uncorrelated case,
takes the form
$$
g(\vek{\eta})=g(\vek{m})+
\nabla g(\vek{m})\cdot(\vek{\eta}-\vek{m})+
\frac12(\vek{\eta}-\vek{m})^T\mathsf{V}(\vek{m})
(\vek{\eta}-\vek{m})+\dots.
$$
Here $\mathsf{V}(\vek{m})$ is the matrix of second derivatives
of the function $g(\vek{\eta})$, calculated at the point
$\vek{\eta}=\vek{m}$. Now we can proceed in the same way as
before
\begin{eqnarray*}
\bavg{g(\vek{\eta})}&\approx&
g(\vek{m})\int f(\vek{\eta})\,\dd\vek{\eta}+
\sum_{i=1}^N\partial_i g(\vek{m})
\int(\eta_i-m_i)f(\vek{\eta})\,\dd\vek{\eta}+\\
&\phantom{=}&+\frac12\sum_{i,j=1}^NV_{ij}
\int(\eta_i-m_i)(\eta_j-m_j)f(\vek{\eta})\,\dd\vek{\eta}=\\
&=&
g(\vek{m})+\frac12
\sum_{i,j=1}^N S_{ij}V_{ij}=
g(\vek{m})+\frac12\,\mathrm{Tr}(\mathsf{S V}).
\end{eqnarray*}
In the last line we used the symmetry of $\mathsf{S}$. For given
$g(\vek{\eta})$, $\vek{m}$ and $\mathsf{S}$, we can now solve
the equation $\avg{g(\vek{\eta})}=0$. In particular, these
approximations can be cast into Eq.~\req{max-condition}, which
can then be treated as in the uncorrelated case.

\end{document}